\documentclass[12pt]{iopart}
\usepackage{iopams}  
\usepackage{color}
\usepackage[normalem]{ulem}
\newcommand{\lambdaC}{{\mkern0.75mu\mathchar '26\mkern -9.75mu\lambda}}

\begin{document}

\title[Atomic Collapse in Graphene: Lost of Unitarity]{Atomic Collapse in Graphene: Lost of  Unitarity}

\author{D~Valenzuela$^1$, S~Hern\'andez-Ortiz$^2$, M. Loewe$^{1,3, 4}$ and A Raya$^2$}

\address{$^1$Instituto de F\'isica, Pontificia Universidad Cat\'olica de Chile, Casilla 306, San\-tia\-go 22, Chile\\
$^2$Instituto de F\'isica y Matem\'aticas, Universidad Michoacana de San Nicol\'as de
Hidalgo, Edificio C-3, Ciudad Universitaria, 58040 Morelia, Michoac\'an, Mexico\\
$^3$Centre for Theoretical and Mathematical Physics and Department of Physics,
University of Cape Town, Rondebosch 7700, South Africa\\
$^4$ Centro Cient\'{\i}fico-Tecnol\'ogico de Valpara\'{\i}so, Casilla 110-V, Valpara\'{\i}so, Chile}
\ead{devalenz@uc.cl, sortiz@ifm.umich.mx, mloewe@fis.puc.cl and raya@ifm.umich.mx}
\vspace{10pt}

\begin{abstract}
We explore the problem of atomic collapse in graphene by monopole impurities, both electric and magnetic, within the context of supersymmetric quantum mechanics. For electric impurities, upon factorizing the radial Dirac Hamiltonian and identifying the supercharges, existence of a critical charge that makes the ground state {\em fall-into-the-center} translates into lost of hermicity for the corresponding Hamiltonian  and hence lost of unitarity of the theory. For the problem of magnetic monopole impurities, preservation of unitarity for all values of the parameters of the corresponding potential translates into the absence of atomic collapse in this case.
\end{abstract}


\vspace{2pc}

\submitto{\JPA}
\maketitle

\section{Introduction}

\noindent An interesting prediction of relativistic quantum mechanics is the possibility of having processes whereby the neutral vacuum of quantum electrodynamics (QED) becomes unstable, inducing  emission of positrons~(see, for instance, Ref.~\cite{Greiner} and references therein). In this way the normal  neutral vacuum decays into a charged vacuum. This process, in spite of many efforts, however, has not yet been observed because it requires a Coulomb potential generated by a nucleus with a charge bigger than $Z=137$. The collapse of the neutral vacuum is triggered by the diving of bound states into the Dirac sea, the negative energy continuum states. If such a situation occurs, the dived state becomes degenerate with the occupied negative energy electron states of the sea and one of these electrons may occupy the new state, leaving a hole, i.e. a positron, that escapes while the electron remains near the source of the potential. 

Going into a completely new scenario, the physics of graphene an other Dirac-Weyl type of materials, where a linear ``relativistic'' dispersion relations appears at low energies~(see Ref.~\cite{vafek} for a recent review), we may use this picture to discuss, for example, the huge peaks in longitudinal conductivity of graphene doped with external impurities as calcium atoms~\cite{experiment}. More recently, it has been observed that a single-atom vacancy in graphene can host a
local charge, which can be gradually built up by applying voltage pulses with the tip of a scanning tunnelling microscope till it reaches a supercritical regime~\cite{experiment2}. This happens because in these materials, particularly graphene, where electrons move in a two-dimensional manifold, the scale is set by the Fermi velocity $v_F$ which is around 300 times slower than the speed of light in vacuum $c$. Hence, we expect the ``relativistic'' effects  to be enhanced by the same factor.

The problem of atomic collapse by electric impurities in graphene  has been discussed by many authors in the past~\cite{otros,katsnelson,castroneto,gusynin}. The setup of the problem is the (2+1)-dimensional Dirac equation in a Coulomb potential $V(r)\sim -1/r$. Bound states are found and then, the key observation is that the energy of the ground state becomes a purely imaginary number when a critical value for the charge of the impurity is reached. Atomic collapse has been discussed for the Dirac equation for the case of short-range potentials in the form of $\delta$-functions~\cite{marcelo,marcelo2}. For a 3-dimensional spherical $\delta$-potential~\cite{marcelo}, conditions for the wave functions to reach the critical state, namely, the parameters which dive the ground state to the continuum are discussed. As opposed to the spherical case, for a cylindrical $\delta$-potential~\cite{marcelo2}, which allows a reduction to (2+1)-dimensions, the critical state cannot be reached due to the lack of support of the potential. In this work, we present a new perspective to the atomic collapse induced by a long-range electric monopole impurity. Using the ideas of supersymmetric quantum mechanics (SUSY-QM), we are able to show, in a quite natural way,  that the atomic collapse is related to lost of unitarity, as expected. In fact, we introduce a rotation operator that allows us to decouple the radial Dirac equation obeyed by the pseudo-spinor in the presence of an external spherically symmetric potential induced by external charges. The emerging system can be cast in the form of a supersymmetric pair. This construction is not possible anymore when the effective charge becomes bigger than a certain critical value which agrees with previous results in the literature. 

We further extend our framework to the case of graphene in a magnetic monopole impurity, first considered in Ref.~\cite{indios}. The Dirac equation in this  case, properly factorized, corresponds to a supersymmetric pair with magnetic Coulomb potentials~\cite{david}. Nevertheless, in this case, atomic collapse is not realized. Such an observation is in agreement with the classical statement that magnetic fields do not exert work on charged particle systems. The remaining of the article is organized as follows: In Sect.~\ref{sec:susy}, we briefly remind the reader the basics of SUSY-QM. Sect.~\ref{sec:impurities} presents a detailed analysis of the problem of graphene in electric and magnetic monopole impurities. We  finally discuss our findings and present our conclusions in Sect.~\ref{sec:conclu}. Further details of the calculation are presented in an appendix.

\section{Brief survey on supersymmetric quantum mechanics}~\label{sec:susy}

Supersymmetric quantum mechanics, or SUSY-QM, is a framework which relates the spectrum and wave functions of two quantum mechanical systems whose Hamiltonians contain the potentials $V_1$ and $V_2$ related in a specific manner we describe below~\cite{review,gamboa}. For a particle of mass $m$ under the influence of the a one-dimensional potential $V_1(x)$, in units where $\hbar=2m=1$, let us consider the Hamiltonian
\begin{equation}
H_1=-\frac{d^2}{dx^2}+V_1(x)\;,
\end{equation}
which can be factorized in the form
\begin{equation}
H_1=A^\dagger(x)A(x)\;,
\end{equation}
where the operators $A(x)$ and $A^\dagger(x)$ are
\begin{equation}
A(x)=\frac{d}{dx}+W(x)\;,\qquad
A^\dagger(x)=-\frac{d}{dx}+W(x)\;.\label{AAdag}
\end{equation}
Here, $W(x)$ is referred to as the super-potential, and is related to the original potential $V_1(x)$ through the Riccati equation
\begin{equation}
V_1(x)=W^2(x)-W'(x)\;,
\end{equation}
where the prime denotes derivative with respect to $x$.  The superpartner of $H_1$ is a second Hamiltonian expressed in terms of a potential $V_2(x)$, 
\begin{eqnarray}
H_2
&=& -\frac{d^2}{dx^2}+V_2(x)\;,
\end{eqnarray}
and is  such  that it can be factorized as 
\begin{eqnarray}
H_2&=&A(x)A^\dagger(x) \;, 
\end{eqnarray}
where the superpartner potential $V_2(x)$ is related to the superpotential $W(x)$ through
\begin{equation}
V_2(x)=W^2(x)+W'(x)\;.
\end{equation}
From the stationary Schr\"odinger equations,
\begin{eqnarray}
H_1\psi^{(1)}_n(x) &=& A^\dagger(x)A(x)\psi^{(1)}_n(x) \ = \ E_n^{(1)} \psi^{(1)}_n(x) \;,\label{Schh1}\\
H_2\psi^{(2)}_n(x) &=& A(x)A^\dagger(x)\psi^{(2)}_n(x) \ = \ E_n^{(2)} \psi^{(2)}_n(x)\label{Schh2} \;,
\end{eqnarray}
we observe that
\begin{equation}
H_2 (A(x)\psi_n^{(1)}(x))= A(x)A^\dagger(x)A(x)\psi_n^{(1)}(x) = E_n^{(1)}(A(x)\psi_n^{(1)}(x))\;,
\end{equation}
and also,
\begin{equation}
H_1 (A^\dagger(x)\psi_n^{(2)}(x))= A^\dagger(x)A(x)A^\dagger(x)\psi_n^{(2)}(x) = E_n^{(2)}(A^\dagger(x)\psi_n^{(2)}(x))\;.
\end{equation}
Assuming the ground state energy $E_0^{(1)}=0$, we have that
\begin{eqnarray}
E_n^{(2)}&=& E_{n+1}^{(1)}\;, \qquad E_0^{(1)}=0\;,\nonumber\\
\psi_n^{(2)}&=& [E_{n+1}^{(1)}]^{-1/2}A(x)\psi_{n+1}^{(1)}(x)\;,\nonumber\\
\psi_{n+1}^{(1)}&=& [E_{n}^{(2)}]^{-1/2}A^\dagger(x)\psi_{n}^{(2)}(x)\;,
\end{eqnarray}
namely, $H_1$ and $H_2$ are isospectral up to one less bound state which $H_2$ does not have~\cite{review}.

The above quantum mechanical pair of Hamiltonians can be treated simultaneously in an operator form. Defining the extended Hamiltonian
\begin{equation}
\mathbb{H}=\left(
\begin{array}{cc} H_1 & 0 \\  0 & H_2\end{array}
 \right) \;,
\end{equation}
and the {\em supercharges}
\begin{equation}
Q(x)=\left( \begin{array}{cc} 0 & 0 \\ A(x) & 0\end{array}
 \right)\;, \qquad
Q^\dagger(x)=\left(\begin{array}{cc} 0 & A^\dagger(x) \\ 0 & 0\end{array}
 \right)\;, \label{supercharge}
\end{equation}
with $A(x)$ and $A^\dagger(x)$ defined in Eq.~(\ref{AAdag}), the supersymmetric pair is defined in a compact form through the algebra
\begin{eqnarray}
\{ Q(x),Q(x)\} = \{ Q^\dagger(x),Q^\dagger(x)\}= 0,\nonumber\\
 \{ Q(x),Q^\dagger(x)\} = \mathbb{H}\;,\nonumber\\
\left[ Q(x),\mathbb{H} \right] \ =\  \left[ Q^\dagger(x),\mathbb{H} \right] = 0\;.\label{algebra}
\end{eqnarray}

For relativistic particles, massless Dirac equation in two dimensions in a uniform magnetic field perpendicular to the plane of motion of electrons is a typical example of a supersymmetric pair for the upper and lower components of the Dirac spinor~\cite{review,saul}.  Introducing the magnetic field trough minimal coupling, $\boldsymbol{p}\to \boldsymbol{p}+\boldsymbol{A}(\boldsymbol{x})$~\footnote{Here, we are considering $e=1$.}, where $\boldsymbol{A}(\boldsymbol{x})$ is the vector potential giving rise to the magnetic field, $\boldsymbol{B}=\nabla\times\boldsymbol{A}(\boldsymbol{x})$,  the  Dirac Hamiltonian is written as
\begin{equation}
H_D = \boldsymbol{\sigma} \cdot (\boldsymbol{p}-e\boldsymbol{A}(\boldsymbol{x}))\;,\label{hdirac}
\end{equation}
where $\boldsymbol{\sigma}=(\sigma_x,\sigma_y,\sigma_z)$, where the $\sigma_i$ are the Pauli matrices, but we always consider $\boldsymbol{p}=(p_x,p_y,0)$ and $\boldsymbol{A}=(A_x,A_y,0)$ . From the property
\begin{equation}
\left(\boldsymbol{ \sigma}\cdot\boldsymbol{a}\right) \left(\boldsymbol{\sigma}\cdot\boldsymbol{b} \right)=\left(\boldsymbol{a}\cdot\boldsymbol{b} \right)I_{2\times 
2}+i\, \boldsymbol{\sigma}\cdot\left( \boldsymbol{a}\times\boldsymbol{b}\right), \label{property}
\end{equation}
we consider the square of the Hamiltonian~(\ref{hdirac})\;, 
\begin{eqnarray}
(H_D)^2 &=& [\boldsymbol{\sigma} \cdot (\boldsymbol{p}-e\boldsymbol{A}(\boldsymbol{x}))]^2\nonumber\\
&=& (p_x+A_x(\boldsymbol{x}))^2+(p_y+A_y(\boldsymbol{x}))^2+\frac{g}{2}(\nabla\times A(\boldsymbol{x}))_z \sigma_z\nonumber\\
&\equiv&H_P\;,
\end{eqnarray}
which is precisely the Pauli Hamiltonian~\cite{review,saul} for a particle with giromagnetic ratio $g=2$. To see it explicitly, we introduce the supercharges
\begin{eqnarray}
Q^1(\boldsymbol{x})&=&\frac{1}{\sqrt{2}} \left[-(p_y+A_y(\boldsymbol{x}))\sigma_x +(p_x+A_x(\boldsymbol{x}))\sigma_y\right]\;,\nonumber\\
Q^2(\boldsymbol{x})&=&\frac{1}{\sqrt{2}} \left[(p_x+A_x(\boldsymbol{x}))\sigma_x +(p_y+A_y(\boldsymbol{x}))\sigma_y\right]\;.
\end{eqnarray}
Then, it is easy to verify that 
\begin{eqnarray}
\{Q^\alpha(\boldsymbol{x}), Q^\beta(\boldsymbol{x})\} = H_P\delta^{\alpha\beta}\;, \qquad [H_P,Q^\alpha(\boldsymbol{x})]=0\;,
\end{eqnarray}
for $\alpha,\beta = 1,2$.
With this framework, we address the problem of Dirac equation with monopole impurities below.

\section{Electric and Magnetic monopole impurities in graphene}~\label{sec:impurities}

We start our discussion of the
Dirac equation for graphene in stationary, but otherwise arbitrary electromagnetic field, namely 
\begin{eqnarray}
\Big(v_{F}{\boldsymbol{\sigma}}\cdot\left(\boldsymbol{p}-e\boldsymbol{A}(\boldsymbol{x})\right)+\Delta v_{F}^2\sigma_z +V(\boldsymbol{x})-E\Big)\Psi\left(\boldsymbol{x}\right) &=&0\;,\label{eq:dirac}
\end{eqnarray}
where $v_F$ is the Fermi velocity and $\Delta$ is the mass gap for charge carriers, $V(\boldsymbol{x})$ represents the scalar potential, $\boldsymbol{A}(\boldsymbol{x})$ the vector potential.   
For general configurations of electromagnetic fields, it is not obvious than the Hamiltonian in Eq.~(\ref{eq:dirac}) can be factorized in a supersymmetric fahsion. It surely shall be the case for the examples we discuss in the remaining of the article.
Considering isotropic potentials, $V(\boldsymbol{x})=V(r)$ and $\boldsymbol{A}(\boldsymbol{x})=A_r(r)\hat{r}+A_{\theta}(r)\hat{\theta}$, where $\hat{r}$ and $\hat{\theta}$ are the unitary vectors of the polar coordinates $(r,\theta)$ of the system, from the relation  
 $\left( \boldsymbol{\sigma}\cdot\hat{r}\right)^2=I_{2\times 2}$ and the property in Eq.~(\ref{property}),
we re-write the Dirac equation as 
\begin{eqnarray}
\Bigg(-i\,\left( \boldsymbol{\sigma}\cdot\hat{r}\right)\Big[\frac{d}{dr}+i\,\frac{A_r(r)}{2\Phi_b}-\sigma_z\left(\frac{L_z}{\hbar r}+\frac{A_{\theta}(r)}{2\Phi_b}  \right)\Big]\nonumber\\
 +\frac{\sigma_z}{n_g\lambdaC} + \frac{V(r)}{v_F\hbar}-\epsilon \Bigg)\Psi\left( r,\theta\right) =0,\label{col3}
\end{eqnarray}
where $L_z$ is the third component of the angular momentum vector $\boldsymbol{L}=\boldsymbol{r}\times\boldsymbol{p}$ and we have introduced the shorthand notation $\epsilon=E/(v_F\hbar)$ for the scaled energy, $n_g=c/v_F\approx 300$ is the inverse refractive index, $\lambdaC=\hbar/(\Delta c)$ is the reduced Compton wavelength and 
$\Phi_b=\hbar/(2e)=\Phi_0/(2\pi)$, where $\Phi_0$ is the quantum fluxon. Expectedly~\cite{librokatsnelson}, the wave function
$\Psi\left( r,\theta\right)$ is an eigenfunction of the operator $L_z\equiv-id/d\theta$, namely
 of the form $\Psi\left( r,\theta\right)=\psi_l(r)e^{il\theta}$,  with eigenvalues $l\in \mathbb{Z} $. Nevertheless, $\boldsymbol{L}$ is not a conserved quantity. The total angular momentum
$\boldsymbol{J}=\boldsymbol{L}+\boldsymbol{S}$ --where $\boldsymbol{S}$ is the pseudospin operator in this context-- is conserved, and $\Psi(r,\theta)$ is an eigenfunction of the operator $J_z$, namely
\begin{equation}
\Psi(r,\theta)\mapsto \Psi_j(r,\theta): \qquad J_z\Psi_j\left( r,\theta\right)=j\hbar\Psi_j\left( r,\theta\right)\;,
\end{equation}
with $j \in \mathbb{Z}$. Thus, we replace $L_z=J_z-S_z=J_z-\frac{\hbar}{2}\sigma_z$ in Eq.~(\ref{col3}) and realize that $l=j\pm\frac{1}{2}$. Then, we can write
\begin{eqnarray}
\Bigg(-i\,\left( \boldsymbol{\sigma}\cdot\hat{r}\right)\Big[\frac{d}{dr}+\frac{1}{2r}+i\,\frac{A_r(r)}{2\Phi_b}-\sigma_z\left(\frac{j}{r}+\frac{A_{\theta}(r)}{2\Phi_b}  \right)\Big] \nonumber\\
+\frac{\sigma_z}{n_g\lambdaC} + \frac{V(r)}{v_F\hbar}-\epsilon\Bigg)\Psi_j\left( r,\theta\right) =0\;.
\label{col4}
\end{eqnarray}
With the aid of the {\em ansatz}
\begin{eqnarray}
\Psi_j\left( r,\theta\right) &=&\frac{1}{\sqrt{r}}\left(\begin{array}{c}
h_j(r)e^{i\left( j-\frac{1}{2}\right) \theta} \\ i\,g_j(r)e^{i\left( j+\frac{1}{2}\right) \theta}
\end{array}\right) \nonumber\\
& =& \frac{1}{\sqrt{r}}\left(\begin{array}{cc}
e^{i\left( j-\frac{1}{2}\right) \theta}&0\\0&e^{i\left( j+\frac{1}{2}\right) \theta}
\end{array}\right)
\left(\begin{array}{c}
h_j(r)\\i\,g_j(r)
\end{array}\right)\nonumber\\
&\equiv& M_j(\theta)\chi_j(r),
\end{eqnarray}
we rewrite Eq.~(\ref{col4}) as
\begin{eqnarray}
\Bigg(-i\,\left( \boldsymbol{\sigma}\cdot\hat{r}\right)\Big[\frac{d}{dr}+i\,\frac{A_r(r)}{2\Phi_b}-\sigma_z\left(\frac{j}{r}+\frac{A_{\theta}(r)}{2\Phi_b}  \right)\Big]\nonumber\\
 +\frac{\sigma_z}{n_g\lambdaC} + \frac{V(r)}{v_F\hbar}-\epsilon\Bigg) M_j(\theta)\chi_j(r)=0\;.\label{col5}
\end{eqnarray}
We have factorized the angular dependence of the wave function in the matrix
 $M_j\left( \theta\right) $ which can equivalently be expressed as
\begin{eqnarray}
M_j\left( \theta\right)= \left(\begin{array}{cc}
e^{i\left( j-\frac{1}{2}\right) \theta}&0\\0&e^{i\left( j+\frac{1}{2}\right) \theta}
\end{array}\right)&=&e^{i\left( jI_{2\times 2}-\frac{1}{2}\sigma_z\right) \theta}\;.\label{prop1}
\end{eqnarray}
Then, it is straightforward to verify that $M_j(\theta)$ fulfills the property
\begin{eqnarray}
\left(\boldsymbol{\sigma}\cdot\hat{r}\right)M_j\left( \theta\right)&=&M_j\left( \theta\right)\sigma_x\;. \label{prop4}
\end{eqnarray}    
which allows us to write
\begin{eqnarray}
 M_j(\theta)\Bigg(-i\,\Big[\frac{d}{dr}+i\,\frac{A_r(r)}{2\Phi_b}+\sigma_z\left(\frac{j}{r}+\frac{A_{\theta}(r)}{2\Phi_b}  \right)\Big]\sigma_x\nonumber\\
 +\frac{\sigma_z}{n_g\lambdaC} + \frac{V(r)}{v_F\hbar}-\epsilon\Bigg)\chi_j(r)=0\;.
\end{eqnarray}
Thus, defining
\begin{equation}
V_r(r)\equiv \frac{A_r(r)}{2\Phi_b}\sigma_x+\frac{V(r)}{v_F\hbar}I_{2\times 2},
\end{equation}
we write the radial Dirac equation for graphene as
\begin{eqnarray}
\Bigg(-i\,\Big[\frac{d}{dr}+\sigma_z\left(\frac{j}{r}+\frac{A_{\theta}(r)}{2\Phi_b}  \right)\Big]\sigma_x +\frac{\sigma_z}{n_g\lambdaC} + V_r(r)-\epsilon\Bigg)\chi_j(r)&=&0\;,
\label{col6}
\end{eqnarray}
Below we consider this equation in the potential produced by electric and magnetic monopole impurities.

\subsection{Electric monopole impurity}

For the electric monopole impurity, we consider the radial Dirac equation~(\ref{col6}) with the Coulomb potential
\begin{equation}
A_r(r)=A_\theta(r)=0,\qquad V_r\left( r\right) =-\frac{Z\alpha_g }{r}\;,
\end{equation}
where $Z\alpha_g$ is the effective charge of the electric impurity and $\alpha_g=\alpha n_g$ is the fine structure constant of graphene.
Then, the radial Dirac equation becomes
\begin{equation}
\underbrace{\left( 
\begin{array}{cc}
\frac{1}{n_g\lambdaC}+\frac{Z\alpha_g }{r}+\epsilon & \frac{j}{r}-\frac{d}{%
dr} \\ 
\frac{j}{r}+\frac{d}{dr} & \frac{1}{n_g\lambdaC}-\frac{Z\alpha_g }{r}-%
\epsilon%
\end{array}%
\right)}_{\displaystyle{\mathbb{A}}} \left( 
\begin{array}{c}
g_j(r) \\ 
h_j(r)%
\end{array}%
\right) 
=0\;.\label{unrotated}
\end{equation}
The $r$-dependence of the diagonal elements of the matrix $\mathbb{A}$ is such that the components $g_j(r)$ and $h_j(r)$ cannot be decoupled; the  $1/r$ term in both elements brings higher inverse powers of $r$ upon derivation. Decoupling of the equations  can be done by properly rotating $\mathbb{A}$ such that the undesired term vanishes from one of the diagonal elements.
Defining the unitary matrix
\begin{equation}
U(\eta )=e^{i\frac{1}{2}\eta \sigma _{y}}\;,\label{transf}
\end{equation}
where the rotation parameter (angle) $\eta$ is defined through
\begin{equation}
\sin \left( \eta \right) =\frac{Z\alpha_g }{j}\;.\label{param}
\end{equation}
Upon defining $\nu^{2}=j^{2}-\left( Z\alpha_g \right) ^{2}$, we observe that
\begin{equation}
U(\eta )\mathbb{A}U^{\dag }\left( \eta \right) =\left( 
\begin{array}{cc}
-\frac{\nu }{n_g\lambdaC j}-\epsilon & \frac{Z\alpha_g }{\lambdaC j}+%
\frac{\nu }{r}-\frac{d}{dr} \\ 
\frac{Z\alpha_g }{\lambdaC j}+\frac{\nu }{r}+\frac{d}{dr} & \frac{\nu
}{n_g\lambdaC j}-\epsilon-2\frac{Z\alpha_g }{r}%
\end{array}%
\right) \;.
\end{equation}
Observe that only one of the diagonal components of the rotated $\mathbb{A}$ has explicit $r$-dependence. Now, assuming that the radial wave function transforms as
\begin{equation}
U(\eta )\left( 
\begin{array}{c}
g_j(r) \\ 
h_j(r)%
\end{array}%
\right) =\left( 
\begin{array}{c}
G_j(r) \\ 
F_j(r)%
\end{array}%
\right)\;,
\end{equation}
we have that the Dirac equation~(\ref{unrotated}) reduces to
\begin{equation}
\left(
\begin{array}{cc}
-\frac{\nu }{n_g\lambdaC j}-\epsilon & \frac{Z\alpha_g }{\lambdaC j}+%
\frac{\nu }{r}-\frac{d}{dr} \\ 
\frac{Z\alpha_g }{\lambdaC j}+\frac{\nu }{r}+\frac{d}{dr} & \frac{\nu 
}{n_g\lambdaC j}-\epsilon-2\frac{Z\alpha_g }{r}%
\end{array}%
\right) \left( 
\begin{array}{c}
G_j(r) \\ 
F_j(r)%
\end{array}%
\right) =0\;.
\end{equation}
The coupled system of equations for the unknowns $F_j(r)$ and $G_j(r)$ can be decoupled in the standard manner, say, obtaining one of the unknowns from one of the equations and inserting it into the other.  Proceeding in this form, the Dirac equation is equivalent to the decoupled system of equations  
\begin{eqnarray}
\left[ -\frac{d^{2}}{dr^{2}}+\left( \frac{\nu }{r}-\frac{Z\alpha_g \epsilon}{\nu }\right) ^{2}-\frac{d}{dr}\left( \frac{\nu }{r}-\frac{Z\alpha_g 
\epsilon}{\nu }\right) \right] G_j(r)&=&\aleph ^{2}G_j(r)\;,\nonumber\\
\left[ -\frac{d^{2}}{dr^{2}}+\left( \frac{\nu }{r}-\frac{Z\alpha_g \epsilon}{\nu }\right) ^{2}
+\frac{d}{dr}\left( \frac{\nu }{r}-\frac{Z\alpha_g \epsilon}{\nu }%
\right) \right] F_j(r)&=&\aleph ^{2}F_j(r)\;,
\label{system}
\end{eqnarray}
with
\begin{eqnarray}
\aleph ^{2}&=&-\left( \frac{Z\alpha_g }{\lambdaC n_g j}\right) ^{2}-\left( \frac{%
\nu }{n_g\lambdaC j}\right) ^{2}+\epsilon^{2}+\left( \frac{Z\alpha_g }{%
\nu }\epsilon\right) ^{2}\nonumber\\
&=&\left( \frac{\epsilon j}{\nu }\right)
^{2}-\left( \frac{1}{n_g\lambdaC }\right) ^{2}\;.
\end{eqnarray}
In this paper, we are interested in the decoupled system of equations in the form~(\ref{system}). Nevertheless, as discussed in the appendix, the energy eigenvalues can be obtained by the boundary conditions obtained regularizing the Coulomb potential near the origin. These energy eigenvalues are of the form
\begin{equation}
\epsilon_{n,j}=\left( \frac{1}{n_g\lambdaC }\right) \left[1+\frac{Z^2\alpha_g^2}{(\nu+n)^2} \right]^{-1/2}\;,
\end{equation}
where $n\in \mathbb{Z}^+$ for $j>0$ and $n\in \mathbb{N}$ otherwise. For the ground state, $n=0$ and $j=1/2$, it is known that there exists a critical value of the charge $Z_{\rm cr}$ such that
\begin{equation}
2Z_{\rm cr} \alpha_g>1\;,
\end{equation}
and thus the ground state is unstable and through the mechanism of {\em fall-into-the-center}, it collapses~\cite{gusynin}. Atomic collapse has also been observed experimentally as peaks on the longitudinal conductivity~\cite{experiment} and enhancement in the local density of states~\cite{experiment2}. Here, we present an alternative view of this phenomenon through the lost of unitarity. For that purpose, we notice that the system of equations~(\ref{system}) has the form of supersymmetric pair. Indeed, defining the superpotential
\begin{equation}
W_\nu(r)=\frac{\nu }{r}-\frac{Z\alpha_g \epsilon}{\nu }\;,\label{superpot}
\end{equation}
we can also define the rising and lowering operators
\begin{equation}
A_\nu(r)=\frac{d}{dr}+W_\nu(r)\;,\qquad A^{\dag }_\nu(r)=-\frac{d}{dr}+W_\nu(r)\;.
\end{equation}
Then, the system of equations is cast in the form
\begin{eqnarray}
A^{\dag }_\nu(r)A_\nu(r)G_j(r)&=&\aleph ^{2}G_j(r)\;,\nonumber\\
A_\nu(r)A^{\dag }_\nu(r)F_j(r)&=&\aleph ^{2}F_j(r)\;,\label{system2}
\end{eqnarray}
where the SUSY-eigenvalue
\begin{equation}
\aleph^2\quad\mapsto \quad\aleph^2_{n,j}=\left(\frac{Z\alpha_g}{n_g\lambdaC}\right)^2 \frac{(\nu+n)^2-\nu^2}{\left[(\nu+\frac{n}{2})^2-\frac{n^2}{4} \right]^2+(Z\alpha_g \nu)^2}\;. 
\end{equation}
For the ground state, $\aleph_{0,1/2}=0$ as expected.
In matrix form, the system of Eqs.~(\ref{system2}) can explicitly be written
\begin{eqnarray}
\left( 
\begin{array}{cc}
A^{\dag }_\nu(r)A_\nu(r) & 0 \\ 
0 & A_\nu(r)A^{\dag }_\nu(r)%
\end{array}%
\right) \left( 
\begin{array}{c}
G_j(r) \\ 
F_j(r)%
\end{array}%
\right) &\equiv& \mathbb{H}_e(r) \left( 
\begin{array}{c}
G_j(r) \\ 
F_j(r)%
\end{array}%
\right) \nonumber\\
& =&\aleph^{2}_{n,j}\left( 
\begin{array}{c}
G_j(r) \\ 
F_j(r)%
\end{array}%
\right) \;.
\end{eqnarray}
Defining the supercharge
\begin{equation}
Q_\nu(r)=\left( 
\begin{array}{cc}
0 & 0 \\     
A_\nu(r) & 0%
\end{array}%
\right) \quad\Rightarrow \quad
Q^\dagger_\nu(r)=\left( 
\begin{array}{cc}
0 & A^\dagger_\nu(r) \\ 
0 & 0%
\end{array}%
\right) \;,
\label{superchargeC}
\end{equation}
we see that the set of operators $\{Q_\nu,Q^\dagger_\nu,\mathbb{H}_e\}$ close the supersymmetric algebra~(\ref{algebra}).

For supercritical states, it is easy to see that $Z_{\rm cr}$ translates to a critical value of  $\nu_c=(1/2)\sqrt{1-(2Z_{cr}\alpha_g)^2}$ (or $\eta_{\rm cr}$ in Eq.~(\ref{transf})) such that   $\nu_c^2$ becomes a negative number and the state collapses. At $Z_{\rm cr}$, the parameter in Eq.~(\ref{param}) for the transformation~(\ref{transf}) is such that 
$\sin(\eta_{\rm cr}),\ \cos(\eta_{\rm cr})>1,$
which can be achieved only  if $\eta_{\rm cr}$ is purely imaginary. Hence, the transformation in Eq.~(\ref{transf}) is no longer unitary. Thus, lost of unitarity  in this language is equivalent to the statement of  atomic collapse. Notice that at criticality, the superpotential in Eq.~(\ref{superpot}) becomes a purely imaginary function.

Next, we discuss the problem of graphene in the potential of a magnetic monopole impurity.

\subsection{Magnetic Coulomb impurities in graphene}

Now we consider the problem of a monopole magnetic impurity in graphene described by the potential
\begin{equation}
A_\theta(r)=\lambda\;.
\end{equation}
This problem has been considered in Refs.~\cite{indios} and~\cite{david} considering a sequence of cylindrical dipolar magnets with thickness varying in a manner such that the resulting magnetic field falls as the inverse of the distance.

Defining $\ell_\lambda=2\Phi_b/\lambda$, the radial Dirac equation for the magnetic Coulomb potential is equivalent to the coupled system of equations
\begin{eqnarray}
\left[ \frac{d}{dr}+\left( \frac{j}{r}-\bigskip \frac{1}{\ell _{\lambda }}%
\right) \right] g\left( r\right) &=&\left( \epsilon-\frac{1}{n_g\lambdaC}%
\right) h\left( r\right) \;,\nonumber\\
\left[ \frac{d}{dr}-\left( \frac{j}{r}-\bigskip \frac{1}{\ell _{\lambda }}%
\right) \right] h\left( r\right) &=&-\left( \epsilon+\frac{1}{n_g\lambdaC}%
\right) g\left( r\right) \;.\label{systemm}
\end{eqnarray}
Defining
\begin{equation}
w^{2}=\epsilon^{2}-\left( \frac{1}{n_g\lambdaC}\right) ^{2}\;,
\end{equation}
the system of Eqs.~(\ref{systemm}) can be straightforwardly decoupled (in other words, the unitary matrix that decouples the system is the identity) and the unknown functions verify
\begin{eqnarray}
\left[ -\frac{d^{2}}{dr^{2}}+\left( \frac{j}{r}-\bigskip \frac{1}{%
\ell _{\lambda }}\right) ^{2}+\frac{d}{dr}\bigskip \left( \frac{j}{r}%
-\bigskip \frac{1}{\ell _{\lambda }}\right) \right] h\left( r\right)
&=&w^{2}h\left( r\right) \nonumber\\
\left[ -\frac{d^{2}}{dr^{2}}+\left( \frac{j}{r}-\bigskip \frac{1}{\ell
_{\lambda }}\right) ^{2}-\frac{d}{dr}\bigskip \left( \frac{j}{r}-\bigskip 
\frac{1}{\ell _{\lambda }}\right) \right] g\left( r\right) &=&w^{2}g\left(
r\right) \;.\label{sysmag}
\end{eqnarray}
The above system of  Eqs.~(\ref{sysmag}) can be cast in the form of confluent hypergeometric equations with energy eigenvalues (see appendix for further details)~\cite{indios,david}
\begin{equation}
\epsilon_{n,m}=\pm \lambda \sqrt{1-\frac{(2m+1)^2}{(2n-1)^2}}\;,\label{indios}
\end{equation}
with $m\in \mathbb{Z}^+$,  $n\in \mathbb{N}$ and $n>m$. No pair of numbers $(m,n)$ can render these eigenvalues to develop an imaginary part and thus, in this case there is no atomic collapse phenomenon. For the purposes of our analysis, we consider the system of Eqs.~(\ref{sysmag}),  which can be reached without advocating for any rotation whatsoever,  and observe it forms a supersymmetric pair by defining the superpotential
\begin{equation}
W_m(r)=\frac{j}{r}-\bigskip \frac{1}{\ell _{\lambda }}\;.
\end{equation}
With this superpotential, defining the lowering and rising operators as in Eq.~(\ref{AAdag}) and the supercharge as in Eq.~(\ref{supercharge}), by defining the corresponding extended Hamiltonian, 
\begin{equation}
\mathbb{H}_m = \left( 
\begin{array}{cc} 
\left[ -\frac{d^{2}}{dr^{2}}+V_1(r)\right] & 0 \\
0 & \left[ -\frac{d^{2}}{dr^{2}}+V_2(r)\right]
\end{array}
\right)\;,
\end{equation}
where
\begin{equation}
V_{1,2}(r)= W_m^2(r)\pm W_m'(r)\;,
\end{equation}
it is easy to verify that $\{Q(r),Q^\dagger(r),\mathbb{H}_m\}$ close the supersymmetric algebra in Eq.~(\ref{algebra}). We emphasize that the extended Hamiltonian $\mathbb{H}_m$ hence defined is always hermitian.

\section{Concluding remarks}~\label{sec:conclu}

In this article we have explored the Dirac equation for graphene under the influence of monopole impurities, both electric and magnetic. In the former case, factorizing the corresponding Hamiltonian  {\em \'a la} SUSY-QM. In the case of the electric impurity, the problem reduces to the well known example of Dirac equation in the Coulomb potential. It has been established by theoretical calculations and experimental measurements that there exist a critical charge $Z_{\rm cr}$ above which there is a fall-into-the center phenomenon associated with the atomic collapse. Here, we have presented a new look at this phenomenon by realizing that the factorization of the corresponding Hamiltonian in this case looses hermicity at the critical charge, and hence for supercritical states, it no longer preserves probability, which is a statement of the lost of unitarity. 
In the SUSY-QM language this manifests through the superpotential which becomes a purely imaginary function.
On the other hand, for the magnetic impurity, regardless factorization, the Hamiltonian is always hermitian. Thus, there is no collapse in this case in agreement with the classical observation that magnetic fields do not exert work, and hence are incapable of dragging the energy of the ground state to the continuum.

\ack
We acknowledge support from CIC-UMSNH (M\'exico) under grant No. 4.22, CONACyT (M\'exico) under grant 256494, FONDECYT (Chile) under Grants 1150847, 
 1130056 and 1150471, and Proyecto Basal (Chile) FB 0821. AR and SHO acknowledge the hospitality of PUC, where part of this work was carried out. AR specially acknowledges VRID-PUC for partial support.

\appendix
\setcounter{section}{1}
\section*{Appendix A}

In this appendix, we find the energy eigenvalues for the bound states in the potential of the electric impurity.
The Coulomb potential should be regularized at the origin. We introduce a distance $R$, which plays the role of a regulator and which eventually will vanish, such that in 
a vicinity of the origin the Coulomb potential is replaced by the constant potential
\begin{equation}
V_r(r)=-\frac{Z\alpha_g }{R}\;,
\end{equation}
and hence
the radial part of the wave function is:
\begin{equation}
\chi _{\rm in}\left( r\right) =A\bar{\omega}\left( 
\begin{array}{c}
rj_{j-1}(\bar{\omega}r) \\ 
i\sqrt{\frac{Z\alpha_g +\epsilon R-\frac{R}{n_g\lambdaC}}{Z\alpha_g +%
\epsilon R+\frac{R}{n_g \lambdaC}}}rj_{j}(\bar{\omega}r)%
\end{array}%
\right) \equiv \left( 
\begin{array}{c} h_{\rm in}(r) \\ g_{\rm in}(r) \end{array}\right)\;, \label{origin}
\end{equation}
where 
\begin{equation}
\bar{\omega}^{2}\equiv \left( \frac{Z\alpha_g }{R}+\epsilon
\right) ^{2}-\left( \frac{1}{n\lambdaC}\right) ^{2}\;,
\end{equation}
and $j_l(z)$ are the spherical Bessel functions~\cite{tablas}. The above pseudospinor should match the corresponding away from the impurity.

For $r>R$, our starting point is the system of Eqs.~(\ref{system}), 
which can be cast in the form of Whittaker differential equations
\begin{eqnarray}
\left[ \frac{d^{2}}{dx^{2}}+\frac{\frac{1}{4}-\left( \nu -\frac{1}{2}%
\right) ^{2}}{x^{2}}+\frac{\frac{Z\alpha_g \epsilon }{\omega}}{x}-\frac{1}{4}%
\right] F&=&0\;,\nonumber\\
\left[ \frac{d^{2}}{dx^{2}}+\frac{\frac{1}{4}-\left( \nu +\frac{1}{%
2}\right) ^{2}}{x^{2}}+\frac{\frac{Z\alpha_g \epsilon }{\omega}}{x}-\frac{1}{4}%
\right] G&=&0\;,
\label{sysApp}
\end{eqnarray}
where $x\equiv 2wr$ and
\begin{equation}
\omega^{2}\equiv \left( \frac{1}{n_g\lambdaC}\right) ^{2}-\epsilon ^{2}\;.
\end{equation}
Regular solutions to the system~(\ref{sysApp}) at infinity are  
\begin{equation}
F(x)=AW_{\frac{Z\alpha_g \epsilon }{w},\nu -\frac{1}{2}}\left( x\right)\;,\qquad
\bigskip G(x)=BW_{\frac{Z\alpha_g \epsilon }{w},\nu +\frac{1}{2}}\left(
x\right)\;,
\end{equation}
where $A$ and $B$ are normalization constants for the  Whittaker functions $W_{a,b}(z)$. From these solutions, the pseudospinor away from the impurity has the form
\begin{equation}
\chi _{\rm out}(x)=U^{\dag }\left( \eta \right) \left( 
\begin{array}{c}
G\left( x\right)  \\ 
F\left( x\right) 
\end{array}%
\right)\equiv \left( 
\begin{array}{c}
h_{\rm out}\left( x\right)  \\ 
g_{\rm out}\left( x\right) 
\end{array}%
\right) \;.
\end{equation}
Then, boundary condition $\chi _{\rm in}\left( R\right) =\chi _{\rm out}\left( R\right)$ implies
\begin{equation}
\frac{
h_{\rm in}(R)}{g_{\rm in}(R)}=\frac{h_{\rm out}(R)}{g_{\rm out}(R)}\;.
\end{equation}
Now, from the property~\cite{tablas}
\begin{equation}
j_\nu(z)=\sqrt{\frac{\pi}{2z}}J_{\nu+\frac{1}{2}}(z)\;,
\end{equation}
where $J_\nu(z)$ is the Bessel function of the first kind, we observe that the ratio
\begin{equation}
\lim_{R\to 0}\frac{h_{\rm in}(R)}{g_{\rm in}(R)}\equiv\frac{J_{j+\frac{1}{2}}(Z\alpha_g )}
{J_{j-\frac{1}{2}}(Z\alpha_g )}=C_{\rm in}\;,
\end{equation}
with $C_{\rm in}$ a constant independent of $R$. Similarly
\begin{equation}
\lim_{x\to 0} \frac{W_{\frac{Z\alpha_g \epsilon }{\omega},\nu -%
\frac{1}{2}}\left( x\right) }{W_{\frac{Z\alpha_g \epsilon }{\omega},\nu +\frac{1}{2%
}}\left( x\right) }=C\;,
\end{equation}
$C$ being a constant independent of $x$.
Next, we use the asymptotic form
\begin{equation}
W_{\mu ,\gamma }(x)\simeq \frac{\Gamma \left( 2\gamma \right) }{\Gamma
\left( \frac{1}{2}-\mu +\gamma \right) }x^{\frac{1}{2}-\gamma }+\frac{\Gamma
\left( -2\gamma \right) }{\Gamma \left( \frac{1}{2}-\mu -\gamma \right) }x^{%
\frac{1}{2}+\gamma }
\end{equation}
for $x\to 0$. Hence,
\begin{equation}
\frac{W_{\frac{Z\alpha_g \epsilon }{\omega},\nu -\frac{1}{2}}\left(
x\right) }{W_{\frac{Z\alpha_g \epsilon }{\omega},\nu +\frac{1}{2}}\left( x\right) }%
=\frac{\frac{\Gamma \left( 2\nu \gamma -1\right) }{\Gamma \left( -\frac{%
Z\alpha_g \epsilon }{\omega}+\nu \right) }x^{1-\nu }+\frac{\Gamma \left( -2\nu
+1\right) }{\Gamma \left( 1-\frac{Z\alpha_g \epsilon }{\omega}-\nu \right) }x^{\nu
}}{\frac{\Gamma \left( 2\nu +1\right) }{\Gamma \left( 1-\frac{Z\alpha_g
\epsilon }{\omega}+\nu \right) }x^{-\nu }+\frac{\Gamma \left( -2\nu -1\right) }{%
\Gamma \left( -\frac{Z\alpha_g \epsilon }{\omega}-\nu \right) }x^{1+\nu }}\;.
\end{equation}
Defining
\begin{eqnarray}
q(x)&=&\frac{\Gamma \left( -2\nu \right) }{\Gamma \left( 2\nu \right) }\frac{%
\Gamma \left( -\frac{Z\alpha_g \epsilon }{\omega}+\nu \right) }{\Gamma \left( -%
\frac{Z\alpha_g \epsilon }{\omega}-\nu \right) }x^{2\nu }
=\frac{\frac{2\nu C}{-\frac{Z\alpha_g \epsilon }{\omega}+\nu }-\frac{x}{2\nu }%
}{\frac{2\nu }{-\frac{Z\alpha_g \epsilon }{\omega}+\nu }+\frac{Cx}{2\nu }}\;,
\end{eqnarray}
we observe that
\begin{eqnarray}
\lim_{x\to 0} q(x)&=&\lim_{x\to 0} \frac{%
\Gamma \left( -2\nu \right) }{\Gamma \left( 2\nu \right) }\frac{\Gamma
\left( -\frac{Z\alpha_g \epsilon }{\omega}+\nu \right) }{\Gamma \left( -\frac{%
Z\alpha_g \epsilon }{\omega}-\nu \right) }x^{2\nu }\nonumber\\
&=&C\frac{1}{\nu ^{2}-\left( 
\frac{Z\alpha_g \epsilon }{\omega}\right) ^{2}}\equiv\breve{C}\;,
\end{eqnarray}
Therefore, $\Gamma \left( -Z\alpha_g \epsilon /\omega+\nu
\right) $, has simple poles for
\begin{equation}
\frac{Z\alpha_g \epsilon }{\omega}+\nu =-n\qquad n\in \mathbb{N}\;.
\end{equation}
So, finally
\begin{equation}
\epsilon _{n,j}=\frac{1}{n_g\lambdaC j}\left[ 1+\left( \frac{Z\alpha_g }{%
n+\nu }\right) ^{2}\right] ^{-\frac{1}{2}}\;.
\end{equation}

\appendix
\setcounter{section}{2}
\section*{Appendix B}

In this appendix we obtain the energy eigenvalues for the magnetic monopole impurity. Starting from the system of Eqs.~(\ref{sysmag}) we rewrite these equations in a Whittaker form as
\begin{eqnarray}
\left[ \frac{d^{2}}{dx^{2}}+\frac{\frac{1}{4}-\left( j-\frac{1}{2}\right)
^{2}}{x^{2}}+\frac{\frac{j}{k\ell _{\lambda }}}{x}-\frac{1}{4}\right]
h\left( x\right) &=&0\;,\nonumber\\
\left[ \frac{d^{2}}{dx^{2}}+\frac{\frac{1}{4}-\left( j+\frac{1}{2}\right)
^{2}}{x^{2}}+\frac{\frac{j}{k\ell _{\lambda }}}{x}-\frac{1}{4}\right]
g\left( x\right) &=&0\;,\label{sysAB}
\end{eqnarray}
where \
\begin{equation}
x\equiv 2kr\;,\qquad k^{2}=\left( \frac{1}{n_g\lambdaC}\right)
^{2}+\left( \frac{1}{\ell _{\lambda }}\right) ^{2}-\epsilon ^{2}\;.
\end{equation}
Solutions to the system~(\ref{sysAB}) are
\begin{eqnarray}
h\left( x\right)&=&C_{1}W_{\frac{j}{k\ell _{\lambda }},j-\frac{1}{2}}\left(
x\right) \;,\nonumber\\
g\left( x\right) &=&-\frac{k C_{1}}{\epsilon+\frac{1}{n_g\lambdaC}}%
\left( \frac{1}{k\ell _{\lambda }}-1\right) W_{\frac{j}{k\ell _{\lambda }},j+%
\frac{1}{2}}\left( x\right) \;,
\end{eqnarray} 
where $C_{1}$ is the normalization constant. Again, we regularize the magnetic Coulomb potential and consider the constant potential solution around the origin in Eq.~(\ref{origin}). A similar reasoning to the previous appendix reveals that
\begin{equation}
\lim_{x\rightarrow 0} \frac{W_{\frac{j}{k\ell
_{\lambda }},j+\frac{1}{2}}\left( x\right) }{W_{\frac{j}{k\ell _{\lambda
}},j-\frac{1}{2}}\left( x\right) }=C_m\;,
\end{equation}
with $C_m$ a constant independent of $R$. Then, defining
\begin{equation}
q(x)=\frac{\Gamma \left( -2j\right) }{\Gamma \left( 2j\right) }\frac{\Gamma
\left( -\frac{j}{k\ell _{\lambda }}+j\right) }{\Gamma \left( -\frac{j}{k\ell
_{\lambda }}-j\right) }x^{2j}\;,
\end{equation}
we find, after some algebra that
\begin{eqnarray}
\lim_{x\rightarrow 0} q(x)&=&
\lim_{x\rightarrow 0} \frac{\Gamma \left( -2j\right) }{\Gamma \left( 2j\right) }\frac{\Gamma
\left( -\frac{j}{k\ell _{\lambda }}+j\right) }{\Gamma \left( -\frac{j}{k\ell
_{\lambda }}-j\right) }x^{2j}\nonumber\\
&=&\tilde{C}\;,
\end{eqnarray}
where $\tilde{C}$ is a constant and  $k\ell _{\lambda }\neq 1$.
Therefore, $\Gamma \left( -\frac{j}{k\ell _{\lambda }}+j\right) $ has simple
pole in its argument for 
\begin{equation}
-j\left[ \frac{1}{k\ell _{\lambda }}-1\right] =-\breve{n}\qquad \breve{n}\in 
\mathbb{N}\;.
\end{equation}
So, $\forall\ c,j>0$
\begin{equation}
\epsilon _{\breve{n},j}=\sqrt{\left( \frac{1}{n_g\lambdaC}\right) ^{2}-\frac{%
\left( \frac{1}{\frac{\breve{n}}{j}+1}\right) ^{2}-1}{\ell _{\lambda }^{2}}}\;.
\end{equation}
Finally, replacing $\breve{n}=n-(j+1/2)$ we arrive to the desired result, Eq.~(\ref{indios}).

\end{document}